\documentstyle[11pt,fleqn]{article}
\newcommand{\ba}{\begin{eqnarray}}
\newcommand{\ea}{\end{eqnarray}}
\begin{document}
\begin{flushright}
Freiburg--THEP 94/28\\
October 1994\\
\end{flushright}
\vspace{1.5cm}

\begin{center}
{\LARGE\bf Can gravity play a role at the electroweak scale?\footnote{Based on
talks
at the DPG meeting, Dortmund, 1-4 March, 1994 and Bad Honnef, 7-10 march,
1994.}}\\[1cm]
\rm
{\large J.J. van der Bij}\\[.5cm]

{\em Albert--Ludwigs--Universit\"{a}t Freiburg,
           Fakult\"{a}t f\"{u}r Physik}\\
      {\em Hermann--Herder Str.3, 79104 Freiburg}\\[1.5cm]

\end{center}

\bigskip
\begin{abstract}
A connection is made between a model for strongly interacting vector bosons
and the spontaneously broken theory of gravity. The theory contains effectively
no Higgs particle, but should have strong interactions at the electroweak
scale. Some speculations about the nature of these interactions and possible
experimental signatures are discussed.

\end{abstract}
\bigskip
\bigskip
\bigskip
\bigskip

The standard model for the weak interactions describes the presently existing
data well. However, whereas the gauge-structure of the model
has a simple geometrical interpretation, the Higgs part of the model
is not particularly attractive. The Higgs sector is responsible for the
existence of a large number of ununderstood parameters in the theory.
Therefore it is a natural question to wonder whether the Higgs sector is
fundamental. The existence of a fundamental Higgs sector is made even more
questionable because of the so called naturalness problem.

The naturalness problem is the situation that the Higgs mass is quadratically
divergent, after one takes radiative corrections into account. Therefore an
ordinary scale
for the Higgs mass would be the cut-off scale of the theory. However the Higgs
mass is supposed to be of the order of the weak scale. Therefore a fine-tuning
is necessary. Other questions involve the cosmological constant, implying a
possible relation with gravity and the existence of a Landau pole where
the theory breaks down. Altogether this has led to a number of proposals
to eliminate or alter the Higgs sector of the theory.

One example is supersymmetry, which avoids the naturalness problem but leaves
the other problems untouched [1].

 Another way is technicolor where all interactions are
gauge interactions and the symmetry breakdown appears spontaneously. However
no realistic model has been constructed [2].

 A third alternative, to cancel the
quadratic divergences within the standard model itself [3], led to the related
idea of topquark condensates [4]. Also here no realistic model exists.

This leads to the fourth logical alternative, that the cut-off of the theory is
indeed at the weak scale and that strong interactions among the vector bosons
should exist.  Experimentally little is known about the selfinteractions among
the vector bosons. At $\bar pp$ colliders direct limits on the vectorboson
anomalous magnetic moment and quadrupole electric moment have been reported
[5].
Indirect limits on the anomalous couplings via radiative corrections as
measured at LEP are cut-off dependent and do not constrain these couplings
severely. Even if three-vector bosons couplings are absent, this is not a very
severe constraint, since strong interactions are in first approximation only to
be expected at the level of the four vectorboson couplings. This is because
only here one starts to become sensitive to the coupling among longitudinally
polarized vector bosons, which by the equivalence theorem correspond to the
Goldstone bosons of the theory. These Goldstone bosons form a direct probe
of the Higgs sector of the theory. In practice it has therefore been difficult
to construct models with strong interactions in the three vectorboson sector,
whereas possibilities are present in the four vectorboson sector.

A case at hand is the class of models where one introduces extra fields,
having strong interactions which via radiative corrections feed down to the
vectorbosons themselves [6,7].
The example that is to be discussed here is the
model of ref[6]. This is in many ways the simplest extension of the
standard model, containing only one extra singlet, coupling to the Higgs
sector of the theory. The Lagrangian is given by:
$$
{\cal L} = -\frac{1}{2}(D_{\mu} \Phi)^{\dagger}(D^{\mu} \Phi)
-\frac {1}{2}(\partial_{\mu} x)^2 - \frac {\lambda_1}{8}
(\Phi^{\dagger} \Phi -f_1^2)^2  -
\frac {\lambda_2}{8}(2f_2 x-\Phi^{\dagger}\Phi)^2 + {\cal L}_{gauge}
\eqno(1)$$

The physical content of this theory consists of two Higgs fields that mix
with one another. By a suitable choice of parameters one can generate a large
mass splitting between the fields. When this is done the integration over the
heavy fields leads to an effective Lagrangian giving large deviations
even at lower energies. The condition for this to happen is that there
should be a hierarchy of coupling constants in the theory. Otherwise the
decoupling theorem is valid.  In this model the condition is
$\lambda_2 >> \lambda_1 >> 0$.
Ignoring hypercharge, the strong effects can be summarized by the following
effective  Lagrangian :

$$
{\cal L}_{eff} = \alpha_1 Tr(V_{\mu} V^{\mu})Tr(V_{\nu} V^{\nu})
+  \alpha_2 Tr(V_{\mu} V^{\nu})Tr(V^{\mu} V_{\nu})
+ g \alpha_3 Tr( F_{\mu \nu} [V^{\mu},V^{\nu}])
\eqno(2)$$
where
$$
V_{\mu} = (D_{\mu}U)U^{\dagger}
\eqno(3)$$
and
$$
F_{\mu \nu} = (\partial_{\mu} - \frac{ig}{2} \vec W_{\mu} \cdot \vec \tau)
\frac {\vec W_{\nu} \cdot \vec \tau}{2i} -
(\mu \leftrightarrow \nu)
\eqno(4)$$

$U$ is the unitary matrix describing the Goldstone boson fields.
Of particular importance is the parameter $\beta = 128 \pi^2(\alpha_2
-2 \alpha_1)$, which is responsible for the formation of vector resonances.
In the limit $f_2 >> f_1$ one simply has $\beta = \lambda_2/\lambda_1$. This
shows that indeed
$\beta$ can be made arbitrarily large. The presence of the extra interactions
leads in general to resonances in the I=1 sector of the theory [8]. For large
values of $\beta$ the resonances become narrower and lie at lower energy.
Of course the $X$ field here is not to be considered as a fundamental field,
but only as an effective description for an as yet unknown dynamical mechanism.
In ref[6] it was implicitly assumed that the $X$ field had some
relation with technicolor. We will not pursue this connection here, but study
the possibility of a relation with gravity.

 The reasons to assume a connection
between gravity and the Higgs sector are manifold. First there is the
question of the cosmological constant, which is generated by the Higgs
potential. The second reason is that both gravity and the Higgs particle have
some universal characteristics. Gravity couples universally to the
energy-momentum tensor, the Higgs particle to mass, which corresponds to
the trace of the energy-momentum tensor. In the model of ref[6] there is a
further
similarity between the $X$ field and the graviton in the fact that they are
both singlets under the gauge group. An interesting question in gauge theory is
the choice of representations one should take. In the standard model there
exists basically only the fundamental
representation for the fermions and the adjoint for the vector bosons.
Because they have no coupling to ordinary matter, singlet fields are not
well constrained by experiment. Typically one can argue that they are absent
from the theory, because they can have a bare mass term, which can be made
to be of the order of the Planck mass, making these fields invisible.
However one can take the attitude that all masses, including the Planck mass
should be given by spontaneous symmetry breakdown. In this case there is a
hierarchy of mass scales $m_P >> v$. In the spirit of ref[6] we will assume
that this hierarchy is due to a hierarchy in coupling constants and not in
vacuum expectation values of different fields. Given these similarities
it is now natural to consider the $X$ field to be essentially the graviton.
We therefore make the identification $X=c.R$ in the Lagrangian [1],
where $R$ is the curvature scalar.
With this identification the model is a higher derivative theory and as such
not directly useful. We therefore make the low energy expansion ignoring
the higher derivative terms. One is then left with the Lagrangian:

$$ {\cal L} = \sqrt{g}\bigl ( \xi \Phi^{+} \Phi R -\frac {1}{2} g^{\mu \nu}
  (D_{\mu} \Phi)^{+}(D_{\nu}\Phi) -V(\Phi^{+} \Phi) - \frac {1}{4}
F_{\mu \nu} F^{\mu \nu} \bigr )\eqno(5)$$

This is the spontaneous symmetry breaking theory of gravity, with the
standard model Higgs as the origin of the Planck mass. The remnant of the
originally very strong interactions in [1] is the parameter $\xi$,
given by $\xi=1/16 \pi G_N v^2 $.
 This model was recently discussed in [9].
The physical content of the model becomes clear after the Weyl rescaling
 $
g_{\mu \nu} \rightarrow \frac{\kappa^2}{\xi v^2}
g_{\mu \nu}$, giving the Lagrangian :
$$ {\cal L} = \sqrt{g}\bigl ( \kappa^2 R -\frac{3}{2}\frac {\xi v^2}
{\vert \Phi \vert^4}
(\partial_{\mu}
\vert \Phi \vert ^2)(\partial^{\mu} \vert \Phi \vert ^2)
-\frac {1}{2} \frac {v^2}{\vert \Phi \vert^2} (D_{\mu} \Phi^{+})
(D^{\mu} \Phi) - \frac {v^4}{\vert \Phi \vert^4}  V(\vert \Phi \vert^2)
\bigr )
\eqno(6) $$

This theory is basically the standard model without Higgs-particle, as the
Higgs coupling becomes of gravitational strength.
It is therefore non-renormalizable and needs new interactions at the weak
scale. The nature of these interactions is not clear and one can at the moment
only speculate.

One possibility is that at the weak scale strong interactions
are present between different subconstituents of fermions and vector bosons.
The interactions betweeen these subconstituents should not be of the ordinary
gauge type, as technicolor models appear not to work. An example of
fundamentally different interactions could be some form of random dynamics.
The signature of such dynamics at lower energies is not particularly clear.
It appears likely, that some set of pseudo Goldstone bosons could be present.
These pseudo Goldstone bosons would not necessarily form a symmetric manifold,
but chiral dynamics should still be valid. Preliminary investigations [10]
show, that this scenario gives no problems with corrections to the $\rho$
parameter. Constraints from other LEP data are being studied. High
energy colliders should have no problems seeing such pseudo Goldstone bosons.

A second possibility that fits in well with the idea that there is a Higgs
gravity connection is the possibility of the existence of extra dimensions,
the Kaluza-Klein models. For ordinary Kaluza-Klein models towers of states
appear, which can be detected at future colliders. Normally one speaks here of
the TeV scale; I want to emphasize here that such states could appear already
at the weak scale.  Higher dimensions could also show up in a different form.
If the geometry of the extra dimensions plays a role in the dynamics, it is in
principle possible  that extra dimensions are being created in the process of
the collision. These extra dimensions are not necessarily compact. The
signature in this case is  missing overall energy and momentum inside the
detector, but not missing $p_T$. In the design of  detectors one should take
this possibility into account. For hadron colliders this signature would be
rather difficult, as one does not know the energy of the incoming partons.
Only a careful study of distributions could possibly give an answer here.
For high energy
electron-positron machines the situation is much better, while one in principle
knows the energy of the incoming particles. A fully hermetic detector is needed
however.

As a final possibility maybe gravity itself already starts
playing a role at the weak scale. The presence of a zero in the metric
is generally taken to be a place where quantum gravity plays a role.
A zero in the metric corresponds to a zero in the Higgs field here, i.e.
when the energy density of the Higgs field is of the order of 250 GeV.
At first sight the creation of a coherent state  with $\Phi=0$ appears
to be a process of infinitesimal probability. At the tree level one needs
$O(m_P/v)$  Higgs particles each with gravitational coupling strength
to be made. Because the theory is nonrenormalizable however, this may be
a misleading conclusion as there is the very strong coupling $\xi$
present. Therefore higher loop effects could be much more important
than the tree level ones. This is the region of strongly coupled quantum
gravity. A clue to what might happen is given, when one takes the
analogy between [1] and [5] seriously. One has $\beta = O(m_P^2/v^2)$.
This corresponds to vector resonances with a mass of $O(v^2/m_P)$, a
form of composite anti-gravity. Possibly such particles may
play a role in cosmology or in the missing mass problem. To make further
progress along these
lines one should have a formulation of quantum gravity, that allows one
to study strong couplings like $\xi$. As such a formulation is
lacking, presumably one has to look for some form of effective Lagrangian for
gravity that could at least phenomenologically describe the dynamics.
What form such an effective Lagrangian should take is not clear at
present.\\

\noindent {\bf Acknowledgement} I want to thank Prof. M. Consoli for
some interesting discussions.\\

\end{document}